\documentstyle [12pt] {article}
\makeatletter

\@addtoreset{equation}{section}

\ifcase \@ptsize
\or 
\or 
\fi
\advance\textheight by \topskip

\ifcase \@ptsize
\or 
\or 
\fi

\def\@citex[#1]#2{%
\if@filesw \immediate \write \@auxout {\string \citation {#2}}\fi
\@tempcntb\m@ne \let\@h@ld\relax \def\@citea{}%
\@cite{%
  \@for \@citeb:=#2\do {%
    \@ifundefined {b@\@citeb}%
      {\@h@ld\@citea\@tempcntb\m@ne{\bf ?}%
      \@warning {Citation `\@citeb ' on page \thepage \space undefined}}%
      {\@tempcnta\@tempcntb \advance\@tempcnta\@ne%
      \@tempcntb\number\csname b@\@citeb \endcsname \relax%
      \ifnum\@tempcnta=\@tempcntb 
        \ifx\@h@ld\relax%
          \edef \@h@ld{\@citea\csname b@\@citeb\endcsname}%
        \else%
          \edef\@h@ld{\ifmmode{-}\else--\fi\csname b@\@citeb\endcsname}%
        \fi%
      \else
        \@h@ld\@citea\csname b@\@citeb \endcsname%
        \let\@h@ld\relax%
      \fi}%
    \def\@citea{,\penalty\@highpenalty\,}%
  }\@h@ld%
}{#1}}

\makeatother
\begin{document}
\hfuzz=100pt
\textheight 24.0cm
\topmargin -0.5in
%
%
%
%
\newcommand{\be}{\begin{equation}}
\newcommand{\ee}{\end{equation}}
\newcommand{\bea}{\begin{eqnarray}}
\newcommand{\eea}{\end{eqnarray}}
\begin{titlepage}
\makeatletter
\def \thefootnote {\fnsymbol {footnote}} \def \@makefnmark {
\hbox to 0pt{$^{\@thefnmark }$\hss }}
\makeatother
\begin{flushright}
BONN-HE-94-09\\
August, 1994\\
hep-th/9408048
\end{flushright}
\vspace{1.5cm}
\begin{center}
{ \large \bf $N=2$ Current Algebras
             for Non-semi-simple Groups}\\
\vspace{2cm}
{\large\bf Noureddine Mohammedi}
\footnote{e-mail: nouri@avzw02.physik.uni-bonn.de}
\footnote
{Work partially supported by the Alexander von Humboldt-Stiftung.}
\\
\vspace{.5cm}
Physikalisches Institut\\
der Universit\"at Bonn\\
Nussallee 12\\ D-53115 Bonn, Germany\\

\baselineskip 18pt
\vspace{1cm}
{\large\bf Abstract}
\end{center}
We examine the problem of constructing $N=2$ superconformal algebras
out of $N=1$ non-semi-simple affine Lie algebras. These $N=2$
superconformal theories share the property that the super Virasoro
central charge depends only on the dimension of
the Lie algebra. We find, in particular, a construction having a central
charge $c=9$. This provides a possible internal space for string
compactification and where mirror symmetry might be explored.  \\
\setcounter {footnote}{0}
\end{titlepage}
\baselineskip 20pt
\section{Introduction}

The study of $N=2$ superconformal algebras is of fundamental importance
in string theory and topological field theory [1].
It was realised in string
theories that $N=2$ superconformal invariance on the world-sheet
leads to $N=1$ space-time supersymmetry, which is crucial for phenomenological
applications of string theories. To be able to extract physics out of
string theory, the ten-dimensional space-time ${\cal {M}}_{10}$ has to be
compactified as ${\cal {M}}_4\times{\cal {K}}_6$, where ${\cal {M}}_4$
is Minkowski space and ${\cal {K}}_6$ is an internal space which is,
to lowest order in perturbation theory, Ricci flat, complex and
K\"ahler; that is a Calabi-Yau space. However, the physics one reaches
depends very much on which Calabi-Yau manifold one chooses. Therefore
it is natural to look for some criteria which cut down on the number of
the classically  allowed vacua for string theory.
\par
One way of finding this selection procedure among all possible string vacua is
provided by $N=2$ superconformal theories. Indeed, the internal space
${\cal {K}}_6$ can be replaced by any $N=2$ superconformal theory having
a central charge $c=9$. The problem of string compactification reduces then
to studying the space of all possible $N=2$ superconformal field theories.
In this space one moves from one conformal field theory to another (that is,
from one string vacum to another) by perturbing  by the so-called truly
marginal operators in the so-called chiral ring of the $N=2$ superconformal
theory. This chiral ring [2] is at the heart of many interesting features of
$N=2$ superconformal algebras and in particular mirror symmetry [3].
\par
Mirror symmetry relates two Calabi-Yau manifolds ${\cal {K}}_6$ and
${\cal {\widetilde K}}_6$ which are a priori topologically and geometrically
different. If $\left({\cal {K}}_6,{\cal {\widetilde K}}_6\right)$
is a mirror pair
then $h^{(1,1)}_{{\cal {K}}_6}=h^{(2,1)}_{{\cal {\widetilde K}}_6}$
and $h^{(2,1)}_{{\cal {K}}_6}=h^{(1,1)}_{{\cal {\widetilde K}}_6}$,
where $h^{(p,q)}_{{\cal {K}}}$ is a $\left(p,q\right)$-form on ${\cal{K}}$.
Moreover, the non-linear sigma models corresponding to ${\cal {K}}_6$
and ${\cal {\widetilde K}}_6$ must also yield isomorphic $N=2$
superconformal field theories. At the level of the $N=2$
superconformal theory, $h^{(1,1)}_{{\cal {K}}_6}$ and
$h^{(2,1)}_{{\cal {\widetilde K}}_6}$ are identified with the so-called
(antichiral,chiral) truly marginal operators while $h^{(2,1)}_{{\cal {K}}_6}$
and
$h^{(1,1)}_{{\cal {\widetilde K}}_6}$ are identified with the so-called
(chiral,chiral) truly marginal operators [3]. Therefore, one obtains two
different string vacua depending on whether one deforms the
$N=2$ superconformal theory by an (antichiral,chiral) or by a (chiral,chiral)
truly marginal operators. It is crucial to mention that the
(antichiral,chiral) and the (chiral,chiral) truly marginal operators
differ only by the sign of their left $U(1)$ charge.
\par
{}For a geometrical interpretation of mirror symmetry in the context of
Calabi-Yau sigma models one requires that the $N=2$ superconformal
algebra has a central charge $\hat c\equiv 3c$ equal to an integer number.
So far, the only rigorously established  example of mirror symmetry
relies on another model, namely the exactly solvable $N=2$ minimal model [4],
where the mirror manifold is obtained by an orbifold construction [5]. It is,
therefore, crucial to search  for other models where mirror symmetry might be
better understood. It is the purpose of this paper  to provide, as a first
step in this direction, $N=2$ superconformal algebras having integer
conformal anomaly $\hat c$. These algebra are based on
non-semi-simple affine current algebras.  Wess-Zumino-Novikov-Witten (WZNW)
models based on non-semi-simple Lie algebras have recently been extensively
analysed [6--14].
\par
The analogue of our construction for semi-simple affine algebras has
already been performed in [15]. All these $N=2$ constructions are in the
spirit
of Kazama-Suzuki models [16]. Attempts in understanding mirror symmetry
and string compactification in
Kazama-Suzuki models have also been made in [17].
\par
The paper is organised as follows. In section two we review the $N=1$
supersymmetric current algebra based on non-semi-simple Lie algebras.
We also show how this can be cast into a tensor product of a bosonic current
algebra and an algebra of free Majorana fermions. In section three we tackle
the problem of constructing $N=2$ algebras from $N=1$
non-semi-simple affine current algebras. The conditions for the existence of
this
construction are then found. We apply our method to a wide set of Lie algebras
possessing  non-degenerate invariant bilinear forms. Explicit examples
of our construction are also given.

\section{The Supersymmetric Current Algebra}

Let $\cal{G}$ be a Lie algebra (semi-simple or non-semi-simple)
whose commutators are
\be
\left[T^I\,,\,T^J\right]=f^{IJ}_{\,\,\,\,\,\,K}T^K\,\,\,.
\ee
The supercurrent algebra built on this Lie algebra
is given by the operator
product expansions of the supercurrents ${\bf J}^I(Z)$
(left-handed part)
[18]
\be
{\bf {J}}^I(Z_1){\bf {J}}^J(Z_2)=
h^{IJ}Z_{12}^{-1}+f^{IJ}_{\,\,\,\,\,\,K}Z_{12}^{-1/2}
{\bf J}^K(Z_2)\,\,\,.
\ee
In these expressions $Z=(z,\theta)$ denotes the holomorphic
coordinate of two-dimensional superspace and
the symbol $Z_{ij}^M$, $M \in {\bf Z}$,  is defined by
\be
Z_{ij}^M=\left\{\begin{array}{ll}
(z_i-z_j-\theta_i\theta_j)^M\,\,\,, & M\in {\bf Z}\\
(\theta_i-\theta_j)(z_i-z_j-\theta_i\theta_j)^{M-1/2}\,\,\,,
& M\in {\bf Z}+{1\over 2}\end{array}\right.\,\,\,.
\ee
\par
The associativity of the above product expansion shows that
$h^{IJ}$ is an invariant of the group obeying
\be
f^{IJ}_{\,\,\,\,\,\,K}h^{KL}+
f^{IL}_{\,\,\,\,\,\,K}h^{KJ}=0\,\,\,.
\ee
At this stage we do not know if $h^{IJ}$ needs to be invertible.
\par
The super energy-momentum tensor is assumed to take the
general form
\be
{\bf T}(Z)=\Omega_{IJ}:D{\bf J}^I{\bf J}^J:(Z)
+M_{IJK}:{\bf J}^I:{\bf J}^J{\bf J}^K::(Z)\,\,\,,
\ee
where $\Omega_{IJ}$ is symmetric and $M_{IJK}$ is totally antisymmetric.
The super covariant derivatives is $D={\partial\over
{\partial\theta}}+\theta{\partial\over{\partial z}}$ obeying
$D^2=\partial$.
\par
The two tensors $\Omega_{IJ}$ and $M_{IJK}$ are then determined by requiring
that the
supercurrents ${\bf J}^I(Z)$ are primary operators of dimension
${1\over 2}$ with respect to the super stress tensor and that
${\bf T}(Z)$ satisfies the super Virasoro algebra. These requirements
lead uniquely to [10]
\bea
&\Omega_{IK}h^{KJ}={1\over 2}\delta^J_I\,\,\,,\,\,\,
\Omega_{IK}f^{KL}_{\,\,\,\,\,\,J}+\Omega_{JK}f^{KL}_{\,\,\,\,\,\,I}=0&
\nonumber\\
&M_{IJK}={2\over 3}f_{IJK}
\equiv {2\over 3}\Omega_{IP}\Omega_{JQ}
 f^{PQ}_{\,\,\,\,\,\,K}\,\,\,.&
\eea
Therefore $\Omega^{IJ}$ (and $h^{IJ}$) is an invertible
invariant bilinear form of the group ${\cal G}$.
Using these expressions for $M_{IJK}$
and $h^{IJ}$, the
super energy-momentum tensor does indeed satisfy a
super Virasoro algebra with central charge given by [10]
\be
c={3\over 2}\dim({\cal G})-\gamma^{IJ}\Omega_{IJ}\,\,\,.
\ee
In these expression $\gamma^{IJ}$ is the Killing-Cartan invariant
bilinear form defined by
\be
\gamma^{IJ}=f^{KI}_{\,\,\,\,\,\,L}f^{LJ}_{\,\,\,\,\,\,K}\,\,\,
\ee
and is degenerate for non-semi-simple Lie algebras. It was shown in [11]
that for non-semi-simple Lie algebras,
which cannot be decomposed into  a product of semi-simple and non-semi-simple
bits, we
necessarily have $\gamma^{IJ}\Omega_{IJ}=0$. Therefore, the super Virasoro
central charge depends only on the dimension of the Lie algebra.
\par
The supercurrent algebra can be made to look like a direct
sum of a bosonic current algebra and an algebra of free
Majorana fermions. This can be achieved by decomposing ${\bf{J}}^I(Z)$
as
\be
{\bf J}^I(Z)=\psi ^I(z)+\theta J^I(z)
\ee
and introducing a modified
current $\widehat{J}^I(z)$ defined by
\be
\widehat{J}^I(z)=J^I(z)+R^I_{\,\,\,\,JK}:\psi^J\psi^K:(z)\,\,\,\,,
\ee
where $R^I_{\,\,\,\,JK}$ is antisymmetric in the indices
$J$ and $K$, and is determined by requiring that the fermions
$\psi^I(z)$ and the bosonic currents $\widehat {J}^I$
decouple. Indeed, the following operator product expansions
\bea
\psi^I(z_1)\psi^J(z_2)&=&{h^{IJ}\over {(z_1-z_2)}}
\nonumber\\
\widehat{J}^I(z_1)\psi^J(z_2)&=&0
\nonumber\\
\widehat{J}^I(z_1)\widehat{J}^J(z_2)&=&{\left(h^{IJ}-{1\over 2}
\gamma^{IJ}\right)\over (z_1-z_2)^2}
+f^{IJ}_{\,\,\,\,\,\,K}{\widehat{J}^K(z_2)
\over{(z_1-z_2)}}
\eea
hold only if the tensor $R^I_{\,\,\,\,JK}$ satisfies [10]
\be
R^I_{\,\,\,\,JK}=\Omega_{JL}f^{LI}_{\,\,\,\,\,\,K}\,\,\,.
\ee
\par
The super stress tensor is also written in components as
\be
{\bf T}(Z)={1\over 2}G(z)+\theta T(z)\,\,\,\,.
\ee
In terms of the modified currents we have
\bea
T(z)&=&\Omega_{IJ}
:\widehat{J}^I\widehat{J}^J:(z)-
\Omega_{IJ}:\psi^I\partial\psi^J:(z)\nonumber\\
G(z)&=&2\Omega_{IJ}:\psi^I\widehat{J}^J:(z)-{2\over 3}f_{IJK}
:\psi^I:\psi^J\psi^K::(z)\,\,\,\,.
\eea
The components $T(z)$ and $G(z)$ satisfy the usual $N=1$
superconformal algebra. The contribution to the central charge
due to the stress tensor of
the bosonic currents $\widehat{J}^I(z)$ is
$\left[\dim({\cal G})-\gamma^{IJ}\Omega_{IJ}\right]$
while that of the
fermions $\psi^I(z)$ is given by ${1\over 2}\dim({\cal G})$.
These two contributions add up to give, as expected, the central
charge for the $N=1$ algebra calculated in (2.8).

\section{The $N\,=\,2$ Construction}

In order to build the $N\,=2\,$ superconformal current algebra, we
need another $N\,=\,1$ supercurrent. This we write as
\be
\widetilde G(z)=2D_{IJ}:\psi^I\widehat{J}^J:(z)-{2\over 3}S_{IJK}
:\psi^I:\psi^J\psi^K::(z)\,\,\,\,,
\ee
where $S_{IJK}$ is totally antisymmetric.
The quantities $D_{IJ}$ and $S_{IJK}$ are determined by the closure of
the $N\,=\,2$ superconformal algebra.
Let us also define
\be
G^+={1\over\sqrt{2}}\left(G+i\widetilde G\right)\,\,\,\,,\,\,\,\,
G^-={1\over\sqrt{2}}\left(G-i\widetilde G\right)\,\,\,.
\ee
\par
We would like now to demand the closure of the $N\,=\,2$
superconformal algebra. By demanding that
\be
G^{\pm}(z)G^{\pm}(w)=0
\ee
we find a set of equations (spelled out in details in ref.[19])
which are solved by
\bea
&D_{IJ}=-D_{JI}\,\,\,\,,\,\,\,\, D_{IK}\Omega^{KL}D_{LJ}=
-\Omega_{IJ}& \\
&f_{IJK}=
 D_{IP}D_{JQ}f^{PQ}_{\,\,\,\,\,\,K}
+D_{KP}D_{IQ}f^{PQ}_{\,\,\,\,\,\,J}
+D_{JP}D_{KQ}f^{PQ}_{\,\,\,\,\,\,I} & \\
&S_{IJK}=D_{IP}D_{JQ}D_{KL}f^{PQL}&\,\,\,.
\eea
The conditions (3.4) means that $D_{IJ}$ is an almost complex structure
on ${\cal G}$, while equation (3.5) is a constraint on the structure
constants $f^{IJ}_{\,\,\,\,\,\,K}$. The last equation defines, however,
the tensor $S_{IJK}$.
The group indices are raised and lowered using the
invariant bilinear form $\Omega^{IJ}$ and its inverse $\Omega_{IJ}$.
\par
The $U(1)$ current, $H(z)$, necessary for the
existence of the $N\,=\,2$
superconformal algebra is read from the
following operator product expansion
\be
G^+(z)G^-(w)={2c\over {3(z-w)^3}}+{2H(w)\over {(z-w)^2}}
+{2T(w)+\partial H(w)\over {(z-w)}}
\ee
and is found to be
\be
H(z)=i\left[\left(D_{IJ}+D_{PQ}f^{PQ}_{\,\,\,\,\,\,L}
f^L_{\,\,\,\,\,\,IJ}\right):\psi^I\psi^J:(z)
-D_{PQ}f^{PQ}_{\,\,\,\,\,\,I}\widehat J^I(z)\right]\,\,\,.
\ee
The central charge $c$ and the energy-momentum tensor $T(z)$
are those of the previous section.
By virtue of the above conditions, we also have
\be
H(z)G^{\pm}(w)=\pm {G^{\pm}(w)\over {(z-w)}}\,\,\,.
\ee
The operator product expansions in (3.3), (3.7) and (3.9) are
the minimal data necessary to guarantee that we have a
$N\,=\,2$ superconformal algebra [19,20].

\section{$N\,=\,2$ Construction for Lie Algebras with an
Invariant Metric}

Let us begin this section by recalling how one constructs
Lie algebras with invariant and nondegenerate metrics.
This construction unifies all the non-semi-simple Lie
algebras known so far [21,11].
Let $h$ be any Lie
algebra and let $h^*$ denotes its dual. If we choose a basis
$\{H^a\}$ for $h$ then the dual basis for $h^*$ is $\{H_a\}$
such that $<H^a,H_b>=\delta^a_b$.
We also assume that $h$ possesses an invariant bilinear form
(possibly degenerate) which we denote $\omega^{ab}$.
A Lie algebra  structure is defined on the vector space
$h\oplus h^*$ in the following manner. On $h$ one has the
Lie bracket $\left[H^a,H^b\right]=f^{ab}_{\,\,\,\,\,\,c}H^c$,
whereas $h^*$ is made abelian $\left[H_a,H_b\right]=0$.
The mixed brackets are given by $\left[H^a,H_b\right]=
-f^{ac}_{\,\,\,\,\,\,b}H_c$.
\par
In addition to $h$ and $h^*$, we introduce another
Lie algebra $g$ with a nondegenerate invariant metric. This
invariant metric is denoted $\omega^{ij}$ relative to a basis
$\{X^i\}$. The Lie bracket on $g$ is $\left[X^i,X^j\right]=
f^{ij}_{\,\,\,\,\,\,k}X^k$. The action of $h$ on $g$ is given by
antisymmetric derivations [21,11]. Explicitly, this means the existence of
mixed structure constants, $f^{ai}_{\,\,\,\,\,\,\,j}$, satisfying [21,11]
\bea
&f^{ai}_{\,\,\,\,\,\,\,k}\omega^{kj}=-f^{aj}_{\,\,\,\,\,\,k}
\omega^{ki}&\nonumber\\
&f^{ij}_{\,\,\,\,\,\,k}f^{ak}_{\,\,\,\,\,\,l}=
f^{ai}_{\,\,\,\,\,\,k}f^{kj}_{\,\,\,\,\,\,l}+
f^{ik}_{\,\,\,\,\,\,l}f^{aj}_{\,\,\,\,\,\,k}&\,\,\,\,.
\eea
These last relations define on the vector space $g\oplus h\oplus h^*$
the following non-vanishing brackets [21,11]
\bea
\left[X^i,X^j\right]&=&f^{ij}_{\,\,\,\,\,\,k}X^k
+f^{ai}_{\,\,\,\,\,\,k}\omega^{kj}H_a\nonumber\\
\left[H^a,H^b\right]&=&f^{ab}_{\,\,\,\,\,\,c}H^c\nonumber\\
\left[H^a,X^i\right]&=&f^{ai}_{\,\,\,\,\,\,k}X^k  \nonumber\\
\left[H^a,H_b\right]&=&-f^{ac}_{\,\,\,\,\,\,b}H_c\,\,\,.
\eea
This algebra admits an invariant metric given by
\be
\Omega^{IJ} =
 \bordermatrix{& X^j & H^b & H_b \cr
  X^i& \omega^{ij} &  0         &  0        \cr
  H^a&    0   & \omega^{ab}     & \delta^a_b\cr
  H_a&    0   & \delta^b_a &  0        \cr}\,\,\,.
\ee
%
%
This bilinear form is nondegenerate and its inverse is given by
\be
\Omega_{IJ}=\left(
\begin{array}{ccc}
\omega_{ij}&0&0\\
0&0&\delta^a_b\\
0&\delta^b_a&-\omega^{ab}
\end{array}\right)\,\,\,,
\ee
where $\omega_{ij}$ is the inverse of $\omega^{ij}$. Notice
that we have a family of invariant metrics, $\Omega^{IJ}$, parametrised
by $\omega^{ab}$, the invariant bilinear forms of $h$.
\par
In order to apply the formulae of the $N\,=\,2$ construction to the
above Lie algebra, we need to find the corresponding complex structure
$D_{IJ}$. {}For this we suppose that we have a complex structure
$d_{ij}$ on $g$ obeying
\be
d_{ij}=-d_{ji}\,\,\,,\,\,\,d_{ik}\omega^{kl}d_{lj}=-\omega_{ij}
\ee
and that the structure constants of $g$ satisfy
\be
f_{ijn}=
d_{ik}d_{jl}f^{kl}_{\,\,\,\,\,\,\,n}
     +d_{nk}d_{il}f^{kl}_{\,\,\,\,\,\,\,j}
     +d_{jk}d_{nl}f^{kl}_{\,\,\,\,\,\,\,i} \,\,\,.
\ee
These two requirements mean that we have a $N\,=\,2$ superconformal
conformal algebra if we restrict ourselves to the Lie algebra $g$.
\par
We found that the complex structure corresponding to the invariant metric
$\Omega^{IJ}$ is given by
\be
D_{IJ}=\left(
\begin{array}{ccc}
d_{ij}&0&0\\
0&0&i\delta^a_b\\
0&-i\delta^b_a&0
\end{array}\right)\,\,\,.
\ee
Using the structure constants $f^{IJ}_{\,\,\,\,\,\,K}$
of the Lie algebra (4.2) and the expressions for $D_{IJ}$
and $\Omega^{IJ}$, we find that equation (3.5) is satisfied
only if
\bea
f^{ab}_{\,\,\,\,\,\,c}&=&0\nonumber\\
d_{ij}f^{aj}_{\,\,\,\,\,k}&=&d_{kj}f^{aj}_{\,\,\,\,\,\,i}
\,\,\,\,.
\eea
Let us turn our attention to some more concrete examples.

\section{Examples}

As a first example of our construction, let us consider the $n$-dimensional
Heisenberg group whose non-vanishing brackets are given by
\bea
\left[\alpha_i,\alpha^{\dag}_j\right]&=&\delta_{ij}I
\nonumber\\
\left[N,\alpha_i\right]&=&-\alpha_i\nonumber\\
\left[N,\alpha^{\dag}_i\right]&=&\alpha^{\dag}_i \,\,\,,
\eea
where $\{i=1,\dots ,r={{n-2}\over 2}\}$ and $N=\delta^{ij}\alpha_i\alpha^
{\dag}_j$.
The group $g$ in this case is abelian and is generated by
$\left\{X^i=\alpha_1,\alpha^{\dag}_1,\dots ,\alpha_r,\alpha^{\dag}_r\right\}$
while the group $h$ is one-dimensional and is generated by
$\left\{H^a=N\right\}$ and its dual $h^\star$ is generated by
$\left\{H_a=I\right\}$. The invertible invariant bilinear form of this
Lie algebra is [12]
\be
\Omega^{IJ}=\left(
\begin{array}{ccccc}
A_1& &  & & \\
 &\ddots&  & & \\
 & &A_r  & & \\
& & &b& -a\\
& & &-a& 0
\end{array}\right)\,\,\,,\,\,\,
A_1=A_2=\dots =A_r= \left(
\begin{array}{cc}
0&a\\
a&0
\end{array}\right)\,\,\,.
\ee
The corresponding complex structure is found to be
\be
D_{IJ}=\left(
\begin{array}{ccccc}
B_1& &  &  &\\
 &\ddots&  &  &\\
 & & B_r&  &\\
& & &0&{i\over a}\\
& & &-{i\over a}&0
\end{array}\right)\,\,\,,\,\,\,
B_1=B_2=\dots = B_{r}=\left(
\begin{array}{cc}
0&{i\over a}\\
-{i\over a}&0
\end{array}\right)\,\,\,.
\ee
\par
Since $f^{ij}_{\,\,\,\,\,\,k}=0$, equation (4.6) is trivially satisfied.
As required by equation (4.8) we also have $f^{ab}_{\,\,\,\,\,\,c}=0$
and it is staightforward to verify that $d_{ik}f^{ak}_{\,\,\,\,\,\,j}
=d_{jk}f^{ak}_{\,\,\,\,\,\,i}$.
\par
The second example we present here concerns the
centrally extended two-dimensional Euclidean group generated by
$\left\{P_1, P_2, J, T\right\}$ with the commutation relations [6]
\bea
\left[P_i,P_j\right]&=&\epsilon_{ij}T
\nonumber\\
\left[J,P_i\right]&=&\epsilon_{ij}P_j\,\,\,,
\eea
where we identify $\left\{X^i=P_1,P_2\right\}$,
$\left\{H^a=J\right\}$ and
$\left\{H_a=T\right\}$. The invertible invariant bilinear form
$\Omega^{IJ}$ and the complex structure $D_{IJ}$ are given by
\be
\Omega^{IJ}=\left(
\begin{array}{cccc}
a&0&0&0\\
0&a&0&0\\
0&0&b&a\\
0&0&a&0
\end{array}\right)\,\,\,,\,\,\,
D_{IJ}=\left(
\begin{array}{cccc}
0&{1\over a}&0&0\\
-{1\over a}&0&0&0\\
0&0&0&{i\over a}\\
0&0&-{i\over a}&0
\end{array}\right)\,\,\,.
\ee
Here also $f^{ij}_{\,\,\,\,\,\,k}=f^{ab}_{\,\,\,\,\,\,c}=0$ and
$d_{ik}f^{ak}_{\,\,\,\,\,\,j}
=d_{jk}f^{ak}_{\,\,\,\,\,\,i}$.
\par
To conclude , we have explored the possibility of constructing
$N=2$ superconformal algebras out of $N=1$ non-semi-simple
affine Lie algebras. The conditions under which the $N=2$
superconformal algebra exists are spelled out. We present
two non-trivial examples which explicitly solve these
conditions.
\par
The crucial feature of these $N=2$ superconformal algebras
is that they all have integer conformal anomaly $\hat c$.
This would allow for a geometrical interpertation of
the truly marginal perturbations of these $N=2$ superconformal theories.
In particular, when $c=9$ (like in the case of the 6-dimensional
Heisenberg group), the super WZNW model would provide the internal
space for string compactification. The string backgrounds corresponding
to these non-semi-simple groups furnish exact solutions to the
beta functions of the
non-linear sigma models to all orders in perturbation theory [14].
\par
Another important issue regarding this internal space for string
compactification is to find its mirror. {}For this we need to determine the
chiral ring for these $N=2$ superconformal algebras. A possible way
to proceed in determining this chiral ring would be to use a free field
representation of these $N=2$ superconformal theories. This is a natural thing
to do since the central charge is just an integer number.
Finally, it would be also interesting to explore the relation between
our construction and the constructions of Getzler [20] and of
Spindel {{\it et al.}} [22].

\vspace{0.5cm}
\paragraph{Acknowledgements:}
I would like to thank Jos\'e {}Figueroa-O'{}Farrill for discussions
and correspondence. This research
is  partially supported  by the Alexander von
Humboldt-Stiftung.

\end{document}